\documentclass[twocolumn]{emulateapj}  
\usepackage{times}

\usepackage{graphicx,bm,amssymb}
\usepackage{ulem}
\usepackage{epsfig}
\usepackage{amssymb}
\usepackage{natbib}
\usepackage{pstricks}
\usepackage{psfrag}
\usepackage[colorlinks=true,linkcolor=blue,citecolor=blue]{hyperref}
\def\apj{ApJ}
\def\apjl{ApJ}
\def\apjs{ApJS}
\def\aap{A\&A}
\def\mnras{MNRAS}
\def\na{New A}
\def\ssr{Space~Sci.~Rev.}
\def\nat{Nature}
\def\physrep{Phys.~Rep.}

\newcommand{\be}{\begin{equation}}
\newcommand{\ee}{\end{equation}}
\newcommand{\bary}{\begin{eqnarray}}
\newcommand{\eary}{\end{eqnarray}}

\shorttitle{GRB 110721A}
\shortauthors{Fraija N. et al.}

\begin{document}
\title{Modeling the high-energy emission in GRB 110721A and \\ implications on the early multiwavelength and polarimetric observations}

\author{N. Fraija$^{1\dagger}$, W. H. Lee$^1$,  M. Araya$^2$, P. Veres$^3$,  R. Barniol Duran$^4$ and  S. Guiriec$^{5,6}$ } 
\affil{$^1$ Instituto de Astronom\' ia, Universidad Nacional Aut\'onoma de M\'exico, Circuito Exterior, C.U., A. Postal 70-264, 04510 Cd. de M\'exico,  M\'exico.\\
$^2$ Escuela de F\'isica \& Centro de Investigaciones Espaciales (CINESPA), Universidad de Costa Rica, San Jos\'e 2060, Costa Rica\\
$^3$ Center for Space Plasma and Aeronomic Research (CSPAR), University of Alabama in Huntsville, Huntsville, AL 35899, USA\\
$^4$ Department of Physics and Astronomy, California State University, Sacramento, 6000 J Street, Sacramento, CA 95819-6041, USA\\
$^5$ Department of Physics, The George Washington University, 725 21st Street NW, Washington, DC 20052, USA\\
$^6$ NASA Goddard Space Flight Center, Greenbelt, MD 20771, USA
}
\email{$\dagger$nifraija@astro.unam.mx}
%
%
\begin{abstract}
GRB 110721A was detected by the Gamma-ray Burst Monitor and the Large Area Telescope (LAT) onboard the Fermi satellite and the Gamma-ray Burst Polarimeter  onboard the IKAROS solar mission.   Previous analysis done of this burst showed: i) a linear polarization signal with position angle stable ($\phi_p= 160^\circ\pm11$) and high degree of {\small $\Pi=84^{+16}_{-28}$ \%}, ii) an extreme peak energy of a record-breaking at 15$\pm$2 MeV, and iii) a subdominant prompt thermal component observed right after the onset of this burst.    In this paper,  the LAT data around the reported position of GRB 110721A are analysed with the most recent software and then, the LAT light curve above 100 MeV was obtained.  The LAT light curve is modelled in terms of adiabatic early-afterglow external shocks when the outflow propagates into a stellar wind.  Additionally, we discuss the possible origins and also study the implications of the early-afterglow external shocks on the extreme peak energy observed at 15$\pm$2 MeV,  the polarization observations and the subdominant prompt thermal component. 
\end{abstract}


\section{Introduction}

Magnetic field plays an important role in the gamma-ray burst (GRB) physics of relativistic jets \citep[i.e. in their formation, collimation and acceleration,][]{1977MNRAS.179..433B, 2001MNRAS.326L..41K, 2010ApJ...711...50T,2015ASSL..414...45T,2016MNRAS.456.1739B}.  Our understanding of magnetic field properties in the GRB scenery has swiftly increased through the last years due to observational success of polarimetry from prompt and/or early afterglow emission \citep{2009ApJ...695L.208G, 2012ApJ...758L...1Y, 2007Sci...315.1822M, 2009Natur.462..767S, 2015ApJ...813....1K, 2014NewA...29...65P, 2016MNRAS.455.3312G}.
Simulations of magneto-hydrodynamic (MHD) flows have revealed that internal shocks can hardly describe the polarization degree larger than $30$ percent \citep{ 2011ApJ...734...77I}.  A high polarization degree is difficult to reconcile without a magnetic field that is ordered on large scales \citep[e.g. external reverse shocks, ][]{2015SSRv..191..471G}.   
The polarization properties found in some GRBs have shown that outflows can be strongly magnetized and large-scale uniform fields can also survive long after the initial explosion \citep{2013Natur.504..119M, 2003Natur.423..415C,2004MNRAS.350.1288R}.  
Additionally, the values of the magnetic energy density in the reverse-shock region found to be higher than in the forward-shock region have claimed that GRBs are magnetised \citep{2005ApJ...628..315Z, 2003ApJ...595..950Z, 2007ApJ...655..391K, 2012ApJ...755..127S, 2004A&A...424..477F, 2012ApJ...751...33F, 2015ApJ...804..105F, 2017arXiv170509311F}.\\
%
%
GRB 110721A was detected by the Gamma-ray Burst Monitor (GBM) and Large Area Telescope (LAT) onboard Fermi Gamma-ray Space Telescope.  At 04:47:43.75 UT, 2011 July 21, Fermi-GBM triggered on GRB 110721A  \citep{2011GCN..12187...1T} and immediately at 04:47:45, LAT detected high-energy emission from GRB 110721A \citep{2011GCN..12188...1V}.    This burst had an initial peak energy of a record-breaking 15$\pm$2 MeV and a subdominant prompt thermal component, observed right after the onset of the burst \citep{2012ApJ...757L..31A}.  The evolution in time of the extreme peak energy was modeled with a power-law function and the thermal component with a blackbody spectrum.\\ 
%
%
In the photospheric and internal shock scenarios, many authors have discussed the origin of both the subdominant prompt thermal component \citep{2016MNRAS.456.2157I, 2015MNRAS.454L..31A, 2015ApJ...813..127P, 2015ApJ...802..134B, 2013MNRAS.428.2430L} and the brightest peak  \citep{2012ApJ...758L..34Z, 2012ApJ...761L..18V} present in GRB 110721A.  In particular, \cite{2012ApJ...758L..34Z} and \cite{2012ApJ...761L..18V} showed that the brightest peak at 15 MeV cannot be explained in terms of the standard internal shocks scenario but  possibly through the dissipative photospheric synchrotron models in a magnetically-dominated outflow.    \cite{2012ApJ...758L..34Z} proposed that the rapid "hard-to-soft" spectral evolution was consistent with a quick discharge of magnetic energy in a magnetically-dominated outflow. Using the Gamma-ray Burst Polarimeter (GAP) onboard IKAROS solar sail mission, \cite{2012ApJ...758L...1Y} reported polarization measurements in the onset of this burst.  Based on the early measurements done on the polarization degree and position angle, authors suggested that synchrotron emission model was more consistent than photospheric quasi-thermal emission models.\\ 
%
%
In this work, the LAT light curve is obtained and then, it is modelled using the early-afterglow external shock model previously described in \cite{2015ApJ...804..105F} and \cite{2016ApJ...818..190F}.  In addition,  the implications of the early-afterglow external shocks on the extreme peak energy observed at 15 MeV,  the polarization observations and the subdominant prompt thermal component were studied and also discussed.  The paper is arranged as follows: in Section 2 a brief description of the multiwavelength and polarimetric observations is given, and also the results on the analysis of the LAT data are presented.  In Section 3, a successful description of the LAT emission is presented in the context of the early-afterglow external shock model.  In Section 4 we study and discuss the implications of the early-afterglow external shock on the early multiwavelength and polarimetric observations, and brief conclusions are given in Section 5.\\
\section{GRB 110721A}
\subsection{Multiwavelength observations}
GRB 110721A  exhibited a complex spectral and temporal behaviour, similar to LAT  bursts observed by Fermi. This burst  had an initial off-axis angle around $\sim 40^\circ$ in the LAT and the autonomous repointed request triggered by the GBM  brought it down to $\sim 10^\circ$ \citep{2013ApJS..209...11A}.  GRB 110721A triggered the GBM instrument at 04:47:43.75 UT on 2011 July 21  \citep{2011GCN..12187...1T}.  Immediately,  LAT started detecting  high-energy emission since 04:47:45 UT up to more than 200 s \citep{2011GCN..12188...1V}.  Using a standard event selection, LAT observed 20 photons above 100 MeV. The most energetic photon with energy of $6.3 \pm 0.6$ GeV  was detected at 4.5 s after the initial trigger.    The light curve of LAT Low Energy (LATLLE) events displayed a typical  short emission component at early onset.  This short component which lasted much less than the prompt emission,  presented photons between 30 - 130 MeV \citep{2012ApJ...757L..31A}.  The main emission lasting  T$_{90}$= 24.5 s in the GBM, consisted of one considerable episode with a peak flux of $5.7\pm0.2 \times 10^{-5}\, {\rm erg\, cm^{-2}\, s^{-1}}$.   In addition, GRB 110721A  exhibited right after the onset a subdominant prompt thermal component peaking at $\sim 80$ keV and an extreme peak energy ever detected in a burst of  15$\pm$2 MeV \citep{2012ApJ...757L..31A}. \\
From 1840 s to  26 hrs after the trigger,  this burst was observed by the Swift X-ray Telescope (XRT). \cite{2011GCN..12192...1G} and \cite{2011GCN..12212...1G} found a faint X-ray emission from the location of this burst. The Gamma-Ray Burst Optical/Near-Infrared Detector \citep[GROND;][]{2011GCN..12192...1G} and also for more than 3000 sec Gemini Multi-Object Spectrographs \citep[GMOS;][]{2011GCN..12193...1B}  detected its optical counterpart and measured two redshifts:  z=0.382 from the absorption lines of CaII  and z=3.512\footnote{This value is more consistent in accordance with the $E_{p,i} - E_{\gamma,iso}$ correlation \citep{2008MNRAS.391..577A}.} from  Ly-alpha absorption.\\
IKAROS-GAP detected a linear polarization signal with position angle stable ($\phi_p= 160^\circ\pm11$) and high degree of {\small $\Pi=84^{+16}_{-28}$ \%}.\\
\subsection{LAT data analysis}
Event data files were obtained starting at a few seconds before the GBM trigger time for GRB 110721A, 04:47:43 UT on 2011 July 21 \citep{2013ApJS..209...11A}. Fermi-LAT data  above 100 MeV  are reduced using the public database at the Fermi Web site\footnote{\url{http://fermi.gsfc.nasa.gov/ssc/data}}.  These data are analysed with the most recent software {\small \rm SCIENCETOOLS} version v10r0p5\footnote{\url{http://fermi.gsfc.nasa.gov/ssc}} and reprocessed with ``Pass 8'' extended, spacecraft data and the instrument response functions {\small \rm P8R2\_TRANSIENT020\_V6}. Transient events are selected within $15^{\circ}$ of the reported position of the GRB above an energy of 100 MeV with a maximum zenith angle of 100$^{\circ}$. Using the {\small \rm gtfindsrc} Fermi Science tool\footnote{\url{http://fermi.gsfc.nasa.gov/ssc/data/analysis}}  with data obtained within 30 s of the trigger time, the position of the event is found to be at the coordinates (J2000) $\alpha$ = 333$^\circ$.52, $\delta$ = -38$^\circ$.60, with an error circle of radius 0.09$^{\circ}$.\\
Exposure maps are generated with the tool {\small \rm gtexpmap} Fermi Science tool, and standard spectra and response files with the {\small \rm gtbin} and {\small gtrspgen} Fermi Science tools for analysis with the software XSPEC version 12 \citep{1996ASPC..101...17A}. No other sources in the LAT catalog or background emission are considered due to the duration of the event. Data from 0.025 s to the first second after trigger is binned in three logarithmically-spaced time bins and the spectra are fitted with a power-law in each bin. The spectral index is kept frozen to the value obtained for the analysis of the joint data within this time period, 2.6 s.   The resulting fluxes with $1\sigma$ errors in each time bin are calculated after the fit, and then the light curve with the flux above 100 MeV is shown in Figure \ref{fig1} (left panel).  The LAT light curve exhibits a peak at $\sim$ 0.3 s similar to the peak displayed in the LATLLE light curve  between 30  - 130 MeV  \citep{2012ApJ...757L..31A}.
%
\section{Modeling the LAT light curve}%
\begin{figure}
{ \centering
\resizebox*{0.95\textwidth}{0.3\textheight}
{\includegraphics{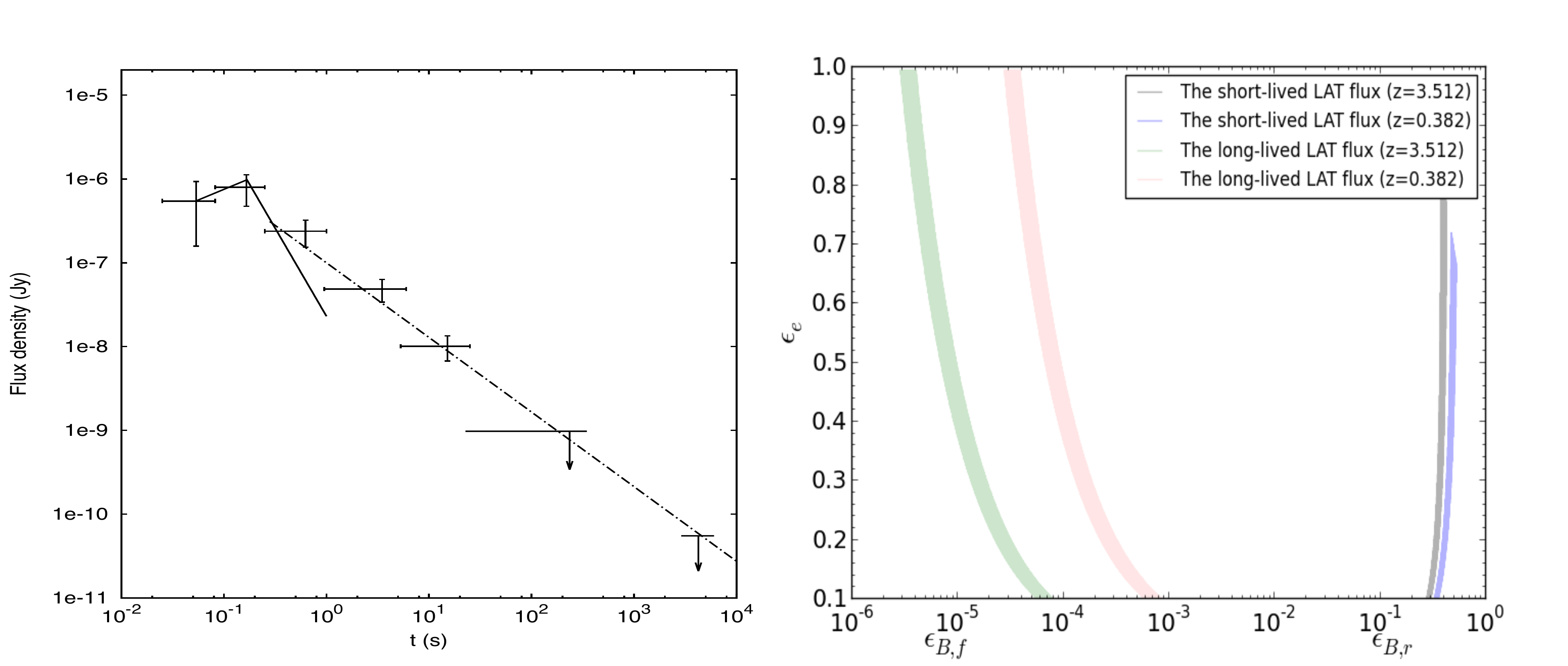}}
}
\caption{Left panel corresponds to the fit of the LAT light curve (above 100 MeV) of GRB 110721A observation with the early-afterglow external-shock model \citep{2015ApJ...804..105F, 2016ApJ...818..190F}. We use the reverse shock in the thick-shell regime to describe the short-lived component (solid line) and the forward shock to explain the long-lived component (dash-dotted line).   For this fit we have used the values of  $E_{\rm \gamma,iso}=1.4\times 10^{52}$ erg and  $E_{\rm \gamma,iso}=1.2\times 10^{54}$ erg for z=0.382 and z=3.512, respectively, and  the values of microphysical parameters obtained in each case are shown in the right panel. This panel shows the values of the microphysical parameters ($\epsilon_{\rm e}$, $\epsilon_{\rm B,f}$ and $\epsilon_{B,r}$), that explain the LAT light curve for z=0.382 and z=3.512.  The parameter region at the left side in this panel reproduces the long-lived LAT flux through synchrotron emission from forward shock and at the right side  describes the short-lived LAT flux by SSC from reverse shock.}
\label{fig1}
\end{figure}

LAT light curve of GRB 110721A presented in Figure \ref{fig1} (left) is similar to the LAT-detected bursts such as GRB090510 \citep{2010ApJ...716.1178A, 2011ApJ...733...22H}, GRB 110731A \citep{2013ApJ...763...71A}, GRB 130427A \citep{2014Sci...343...42A} among other powerful bursts. It has been extensively claimed that GeV photons in the prompt emission could be of early-afterglow external origin \citep[e.g. see][]{2009MNRAS.400L..75K, 2010MNRAS.409..226K, 2010MNRAS.403..926G, 2015ApJ...804..105F, 2016ApJ...831...22F}.   In order to describe the LAT light curve displayed in Section 2, we require the observable quantities of GRB 110721A for the two redshifts z=0.382 and z=3.512 \citep{2011GCN..12193...1B}, and use the early-afterglow external shock model presented in \citet{2015ApJ...804..105F} and  \cite{2016ApJ...818..190F}.  We identify two emission components, one component lasting less than a few of seconds \citep[the so-called short-lived component; see, e.g., ][]{2004A&A...424..477F,  2005ApJ...628..315Z, 2012ApJ...751...33F} and other component lasting more than hundred of seconds \citep[the so-called long-lived component; see e.g.][]{2009MNRAS.400L..75K, 2010MNRAS.409..226K, 2010MNRAS.403..926G, 2011ApJ...733...22H,2016ApJ...831...22F}.  We hereafter use k=$\hbar$=c=1 in natural units, and the values of cosmological parameters $H_0=$ 71 km s$^{-1}$ Mpc$^{-1}$, $\Omega_m=0.27$, $\Omega_\lambda=0.73$  \citep{2003ApJS..148..175S}.   The values of bulk Lorentz factor $\Gamma$ is constrained through the deceleration time  {\small $t_{\rm dec}=9/(64\,\pi)\left(1+z \right)\,\xi^{-2}\,E\,A^{-1}\,\Gamma^{-4}$}, where  $A$ is the stellar wind density,  $\xi$ is a parameter of the order of unity and $E=E_{\gamma,iso}/\eta$ is the isotropic equivalent kinetic energy with $\eta\approx$ 0.2 the kinetic efficiency to convert bulk kinetic energy to $\gamma$-ray energy $E_{\gamma, iso}\simeq 10^{54}$ erg  for z=3.512.   Taking into account the fact that the largest flux density (see the peak flux in fig.\ref{fig1}) was present at $\sim$ (0.25 - 0.35) s, and using the typical values of the electron power index  p = 2.2 and the stellar wind density $A= A_\star (5.0\times 10^{11})\,{\rm g/cm}$ with  $A_\star=0.1$ \citep{2000ApJ...536..195C,  2014Sci...343...42A,2013ApJ...763...71A,  2016ApJ...818..190F,2015ApJ...804..105F}, then the value of the bulk Lorentz factor was found to be $\Gamma\simeq1000$ and the corresponding deceleration radius becomes $r_{\rm dec}\simeq9.9 \times 10^{15}$ cm.  The value of the electron spectral index p = 2.2 was estimated through the slope decays of the LAT long-lived component found $\alpha_{LAT}=1.13\pm0.11$ and the adiabatic synchrotron forward-shock emission in the fast cooling regime $\frac{3p-2}{4}$.   The value of the Lorentz factor found is consistent with the thick-shell regime \citep[$\Gamma_c=410<\Gamma$, with $\Gamma_c$ the critical Lorentz factor; e.g. see][]{2005ApJ...628..315Z, 2003ApJ...595..950Z} for which the ejecta is essentially decelerated by the reverse shock.   The shock crossing time is $t_d\sim\left(\Gamma/\Gamma_c\right)^{-4} T_{90}\simeq 0.5$ s \citep{2007ApJ...655..973K} which is much shorter than the duration of the main burst and it is consistent with the duration of the short-lived component.    It is worth noting that if we would have considered an homogeneous medium in stead of the stellar wind of the progenitor, the value of the bulk Lorentz factor  at the initial deceleration phase of $\sim$  (0.25 - 0.35) s would have been  $\simeq$ 2100 for a typical constant density of n=1 cm$^{-3}$ or $\simeq$ 1000 for n=300 cm$^{-3}$ which in both cases the values are fairly high.\\ 
We did the Chi-square $\chi^2$ minimization using  the ROOT software package \citep{1997NIMPA.389...81B} to fit the LAT light curve shown in fig.\ref{fig1} (left panel) and then, we obtain the values of microphysical parameters for $\xi$= 0.75.  The parameter $\xi$ is a correction factor of the bulk Lorentz factor and the emitting radius due to these quantities are derived to be applied only in the line of sight to the center \citep{1998ApJ...493L..31P, 2000ApJ...536..195C}.   Left panel in Figure \ref{fig1} shows the contributions of synchrotron radiation from forward shock (dash-dotted line) and synchrotron self-Compton (SSC) emission from reverse shock (continuous line) that model the high-energy emission detected in GRB 110721A by the Fermi-LAT instrument, for z=0.382 and z=3.512.   The long-lived high-energy emission was interpreted by synchrotron radiation in the fast cooling regime for relativistic electrons radiating photons around 100 MeV at t= 5 s.  The short-lived high-energy component was fitted with SSC emission from reverse shock in the thick-shell regime when the electron population radiates photons around 100 MeV at t= 0.3 s.    The wind model predicts a rising of t$^{1/2}$ \citep[e.g., see][]{2016ApJ...818..190F} and a steeper decline after the peak of t$^{-5/2}$ \citep{2003ApJ...597..455K} or in some cases t$^{-(p+4)/2}$ when the light curve is determined by the angular time delay effect \citep{2000ApJ...541L..51K}.   It is worth noting that the solutions for the two redshifts are equal for different values of $E_{\rm \gamma,iso}=1.4\times 10^{52}$ erg and  $E_{\rm \gamma,iso}=1.2\times 10^{54}$ erg for z=0.382 and z=3.512, respectively, and  sets of microphysical parameters ($\epsilon_{\rm e}$, $\epsilon_{\rm B,f}$ and $\epsilon_{B,r}$), as shown in Figure \ref{fig1} (right panel). This panel displays the values of microphysical parameters ($\epsilon_{\rm e}$, $\epsilon_{\rm B,f}$ and $\epsilon_{B,r}$), for z=0.382 and z=3.512, that reproduce the short- and long-lived LAT emissions.   The subindexes {\it ``f"} and  {\it ``r"} refer to quantities observed/derived in the forward- and reverse-shock regions, respectively.     As indicated, the parameter space located at the left side in this figure explains the long-lived LAT flux  through synchrotron emission from forward shock and at the right side  describes the short-lived LAT flux by SSC from reverse shock. The values of the microphysical parameters in forward-shock region lie in the ranges of  $0.1\leq\epsilon_{\rm e}\leq 1.0$,  $4\times10^{-6}\leq\epsilon_{\rm B,f}\leq10^{-4}$ ($3\times10^{-5}\leq\epsilon_{\rm B,f}\leq8\times10^{-4}$) for z= 3.512 (0.382) and in reverse-shock region lie in the ranges of  $0.1\leq\epsilon_{\rm e}\leq0.74$  and  $0.32\leq\epsilon_{\rm B,r}\leq0.5$ for z= 0.382   and $0.1\leq\epsilon_{\rm e}\leq0.82$  and  $0.25\leq\epsilon_{\rm B,r}\leq0.35$ for  z= 3.512. \\   
\\
From the range obtained of $\epsilon_{B,r}$,  the magnetization parameter lies in the range of  $0.32\leq\sigma\leq0.5$  and $0.25\leq\sigma\leq0.35$ for z= 0.382  and z=3.512, respectively. The magnetic field derived in forward- and reverse-shock regions can be observed that it is $\simeq$ 50 - 200 (20 - 100) times larger in the reverse-shock region  than in the forward-shock region for z=0.382 (3.512).  The previous results ($\sigma\lesssim$ 1) indicate that the flow is moderately magnetized and then,  a bright $\gamma$-ray flash is expected as presented in some Fermi-detected bursts such as GRB090510 \citep{2010ApJ...716.1178A, 2011ApJ...733...22H, 2016ApJ...831...22F}, GRB 110731A \citep{2013ApJ...763...71A, 2015ApJ...804..105F}, GRB 130427A \citep{2014Sci...343...42A, 2016ApJ...818..190F}.  Otherwise, the reverse shock emission is expected to be weak or suppressed if $\sigma\ll$ 1 or $\sigma\gg$ 1. Even, for a magnetization parameter $\sigma\geq$ 1, the reverse shock emission decreases substantially.\\  
\\
The maximum synchrotron energy at 4.5 s predicts photons with energy of {$E^{\rm syn}_{\rm max,f}  \simeq 4.1\, {\rm GeV}  \left(\frac{1+z}{4.512}\right)^{-3/4}\xi^{-1/2}_{-0.12}E^{1/4}_{54.7}\,A^{-1/4}_{\star,-1}t^{-1/4}_{0.65}$}, then the most energetic photon present of $6.3\pm 0.6$ GeV necessarily must be interpreted by SSC emission from forward shock as explained in some powerful bursts such as GRB 130427A \citep{2013ApJ...776...95F, 2016ApJ...818..190F},  GRB 090926A \citep{2012ApJ...755...12V, 2012ApJ...755..127S}, GRB 09510 \citep{2010ApJ...720.1008C,2016ApJ...831...22F}, see also e.g. \citet{2015MNRAS.454.1073B} and references therein.
\section{Implications of the early-afterglow external shock model  on the early Multiwavelength and polarimetric observations}%
In this section we discuss the possible origins of the extreme peak energy detected at 15 $\pm$ 2 MeV, the polarization observations and the subdominant prompt thermal component.  In addition, we study the implications of the early-afterglow external shock model used to describe the LAT light curve (see subsection 2.2) on these early multiwavelength and polarimetric observations.
 \subsection{The extreme peak energy at 15$\pm$2 MeV}
In the standard fireball model, inhomogeneities in the jet lead to internal shell collisions, higher  shells ($\Gamma_h$) catching slower  shells ($\Gamma_l$).  The kinetic energy of ejecta is partially dissipated via these internal shocks \citep{1994ApJ...430L..93R} which take place at a distance of ${\small r_j=6\times 10^{15}\,{\rm cm}\, \Gamma^2_{\rm is, 3}\, t_{\nu,-1}}$, where $t_v$ is the variability time scale of the central object and  $\Gamma_{\rm is}\simeq\sqrt{\Gamma_h\,\Gamma_l}$ is the bulk Lorentz factor of the propagating internal shocks.   These shocks are expected to be collisionless, so that particles may be accelerated.  The kinetic energy density 
\be
U= \gamma_{\rm sh}/(8\pi)\,\Gamma^{-6}_{\rm is}\,L_j\,t^{-2}_{\nu}\,, 
\ee
with $\gamma_{\rm sh}=\sqrt{\frac{\gamma_{\rm in}}{2}}$ and $L_j$ the isotropic equivalent kinetic luminosity is equipartitioned to accelerate particles $\epsilon_e=U_e/U$ and to generate and/or amplify the magnetic field $\epsilon_B=B^2/(8\pi\,U)$  \citep{2005AIPC..784..164P}  with $B$ the comoving magnetic field given by 
\be
B\simeq \sqrt{\gamma_{\rm sh}}\epsilon_B^{1/2}\Gamma^{-3}_{\rm is}\,L^{1/2}_j\, t^{-1}_\nu\,.
\ee 
Here, $\gamma_{\rm in}$ is the Lorentz factor of internal shock which is of order of a few.  The typical synchrotron energy from internal shock of an electron with a "typical" Lorentz factor ($\gamma_e$) is 
{\small
\be
E^{\rm is}_{\rm p, \langle \gamma_e\rangle} = \frac{e}{m_e}\,\left( 1+z\right)^{-1}\,\Gamma_{\rm is}\,\gamma_e^2\, B.
\ee
}
We consider a  ``typical" electron as one that has the average $\gamma_e$ of the electrons distribution $\langle \gamma_e\rangle=\frac{U_e}{m_e\,N_e}$ \citep{1999PhR...314..575P}. This average value can be estimated in two different cases. In the first case, the energy density carried by electrons and the electron number density are  \citep{1999PhR...314..575P}
{\small
\be
{\rm U_e}=\epsilon_e\,U=\epsilon_e\,\gamma_{\rm sh}\,{\rm N_p\,m_p}\,\,\,\,\, {\rm and} \,\,\,\,\,{\rm N_e}\simeq {\rm N_p}\,. 
\ee
}
Hence,   the average Lorentz factor is $\langle \gamma_e\rangle=\frac{m_p}{m_e}\epsilon_e \gamma_{\rm sh}$.  In the second case,  the energy density carried by electrons and the electron number density are
{\small
\be
{\rm U_e}=\frac{m_e\,A_e}{(p-2)}\gamma^{-p+2}_{\rm e,min}\,\,\,\,\, {\rm and}\,\,\,\,\, {\rm N_e}=\frac{m_e\,A_e}{(p-1)}\gamma^{-p+1}_{\rm e,min},
\ee
}
for $p>2$ and $\gamma_{\rm e,min}\ll\gamma_{\rm e,max}$.  Therefore,   $\langle \gamma_e\rangle=\frac{p-1}{p-2}\,\gamma_{\rm e, min}$.  Summing up both cases,  the observed synchrotron energy becomes
With the typical values, we have
{\small
\bary
E^{\rm is}_{\rm p, \langle \gamma_e\rangle}&\simeq&
\cases{
0.3\,{\rm MeV}\,\,\epsilon^2_{e,-0.3} \gamma^2_{\rm sh}\cr
1.4\,{\rm MeV}\,\,\gamma^2_{\rm e, min,3}\cr
}\cr
&& \hspace{1.1cm}\times \left(\frac{4.512}{1+z}\right)\sqrt{\gamma_{\rm sh}}\epsilon^{1/2}_{B,-1}\,t^{-1}_{\nu,-1}  \Gamma^{-2}_{\rm is, 3} L_{j,52}^{1/2}\,.
\eary
}
It is worth noting that if only a small fraction of the electrons $N_{\rm e, fr}$ are accelerated,  the average Lorentz factor  $\langle \gamma_{\rm e, fr}\rangle=\frac{U_e}{m_e\,N_{\rm e, fr}}$ is larger  than  $\langle \gamma_e \rangle$, and then  $E^{\rm is}_{\rm p, \langle \gamma_{\rm e,fr}\rangle} >E^{\rm is}_{\rm p, \langle \gamma_e\rangle}$ \cite[for a recent discussion see e.g.;][]{2013ApJ...769...69B}.   Including pair formation, a different calculation  indicates an upper limit 
 {\small
  \be
  E^{\rm is}_p\lesssim 0.5\, {\rm MeV}\,  \left(\frac{1+z}{4.512} \right)^{-1}   \, L^{-1/5}_{52} \Gamma^{\frac43}_{\rm is,3}\,t^{\frac16}_{\nu,-1} \epsilon^{\frac12}_{B}\,\epsilon^{\frac43}_{e}\,,
  \ee
  }
which is more rigorous \citep{2001ApJ...557..399G}. Previous analysis indicates that the peak energy can hardy reach values as high as 15 MeV. In addition,  the extreme peak energy of the Band function at 15$\pm$ 2 MeV was measured during the first time bin (from -0.32 to 0.0 s) with a low-energy power law index of $\alpha=-0.81\pm 0.08$ and a high-energy index of $\beta=-3.5^{+0.4}_{-0.6}$ \citep{2012ApJ...757L..31A}. They found that the extreme peak energy decreased monotonically following a power law of the form $E_p= A_{pl}(t-t_0)^\delta$  with  $\delta=-1.22\pm0.13$ for  $t_0=-0.46$ s and $A_{pl}$ the proportionality constant. However,  it has been believed that synchrotron spectrum produced by electrons accelerated in relativistic internal shocks is expected in the fast cooling regime with photon spectral index of  $\alpha=-1.5$ \citep{1986rpa..book.....R}.  It is worth noting that standard internal shocks take place at  $6\times  10^{15}$ cm and the deceleration radius found in this work is $\sim 10^{16}$ cm, then a temporal gap between extreme peak energy and the LAT emission would have been detected, contrary what is observed.  Therefore, the standard internal shocks cannot straightforwardly explain the value of energy peak at 15 $\pm$ 2 MeV,  the power spectral index associated with the initial flux and the timescale observed between the extreme peak energy and the LAT emission described in the external shock framework.\\
\\
%
\cite{2014NatPh..10..351U} showed that considering the effect of adiabatic expansion of magnetic field, the photon index of $\alpha=-0.8$ could be due to synchrotron radiation in the moderately fast cooling regime.  They proposed that a minimum electron Lorentz factor of the order of $10^{5}$ and a strength of the magnetic field in the range of $10 - 100$ G must be presented in the shocks in order to reproduce the value of the photon spectral index of $\alpha=-0.8$.    A feasible scenario to provide  these parameter requirements could be the magnetic dissipation models that use a large dissipation radius, such as the internal collision-induced magnetic reconnection and turbulence (ICMART) events \citep{2011ApJ...726...90Z}.   In the ICMART framework, the Band function is formed at large radii from the central engine, typically at $10^{15} - 10^{16}$ cm.   Electrons are accelerated  by runaway release of the storage magnetic field energy  either in the reconnection zones, or stochastically in the turbulent areas, which emit synchrotron photons that power the prompt emission. During the ICMART event the magnetization parameter decreases fast.  Initially, the magnetization parameter is the order of $\sigma \simeq$ 100, and the discharge process ends when the microphysical parameter is reduced to $\sigma \lesssim 1$.\\
Taking into consideration the two timescales:  the initial time $t_0$=-0.46 of the extreme peak energy at 15$\pm$ 2 MeV  which must have begun before the first bin, \citep[from -0.32 to 0;][]{2012ApJ...757L..31A} and the deceleration time  at $\sim 0.3$ s used to describe the peak of the LAT emission, it is reasonable to  infer that the prompt emission must have taken place at larger radius (10$^{15}$ - 10$^{16}$ cm)  close to the deceleration radius $\sim$10$^{16}$.  On the other hand,  the range of values of the magnetization parameters $\sigma\sim$ 0.3 - 0.5 obtained at the deceleration radius $\sim $10$^{16}$ cm  encourage us to think that before deceleration, the ejecta must also have dissipated a significant amount of Poynting flux during the prompt emission  phase. Therefore, from the analysis performed using the LAT emission can be seen that the most favorable  mechanism  to  make  this  happen and explain in addition the photon index of $\alpha=-0.8$ as synchrotron radiation in the  cooling regime is the ICMART model.   
%
%
%
%
\subsection{Polarization}
%
%
%
Internal shocks may produce strong magnetic fields with random directions on hydrodynamic scales \citep{1999ApJ...511..852G, 2011ApJ...734...77I}.  MHD simulations with initial density fluctuations showed that internal shocks cannot explain the observed polarization degree of $\Pi \gtrsim 30$ \% \citep{ 2011ApJ...734...77I}.  Such is the case of  GRB 100826A, which exhibited  gamma-ray polarization with  polarization angle variable during the prompt emission and polarization degree of $\Pi=27\pm 11$ \%  \citep{2011ApJ...743L..30Y}. Therefore, standard internal shocks could hardly explain the polarization percentage and the behavior of the polarization angle observed in GRB110721A.\\
On the other hand, \cite{2012ApJ...758L...1Y} showed that photospheric emission model cannot describe the polarization degree and the position angle observed in GRB110721A.\\
 \\
The detection of linear polarization in a few percent of GRB afterglows has been accepted as a real confirmation that the synchrotron radiation is the dominant emission mechanism in the afterglow phase.    A high degree of linear polarization with a stable position angle is hard to produce without a magnetic field that is ordered on large scales \citep[e.g. external reverse shocks, ][]{2015SSRv..191..471G}. For  instance, GRB 120308A was observed by the purpose-built RINGO2 polarimeter on the Liverpool Telescope  \citep{2012GCN..13018...1V}.  RINGO2 observations started $\sim$ 240 s after the GBM trigger. Although the position  angle remained almost stable,  the polarization degree showed an evolution from $\Pi=28\pm4$ \% to $\Pi=16^{+5}_{-4}$ \% by 800 s after the GBM trigger \citep{2013Natur.504..119M}.  The analysis performed by \cite{2013Natur.504..119M}  shows that the polarization degree extrapolated to earlier times than $\sim$ 300 s,  must be much higher than $\Pi=28$ \%  as was found in GRB 110721A \citep{2012ApJ...758L...1Y}, thus favouring the early afterglow emission.\\
One of the implications of describing the LAT light curve through the early afterglow emission is that the reverse shock could reproduce the high polarization degree  \citep{2003ApJ...594L..83G}, as  observed in this burst. It is worth nothing that polarization percentage in the emission coming from shocked circumburst medium is expected to be very low \citep[see e.g.;][]{1999A&A...348L...1C, 2003Natur.426..157G}.   Taking into account the energy range observed by IKAROS-GAP (70 - 300 keV), we will show that using the observable quantities of GRB 110721A and the microphysical parameter values found, the synchrotron reverse-shock flux at $E^{\rm syn}=$ 100 keV  dominates over that flux produced in the forward-shock region.   The synchrotron spectral breaks (the characteristic $E^{\rm syn}_{\rm m}$ and the cooling $E^{\rm syn}_{\rm c}$  break energy)  and the maximum synchrotron flux using z=3.512, for the forward and reverse shocks are \cite[see e.g.,][]{2015ApJ...804..105F}
{\small
\bary\label{syn_for}
E^{\rm syn}_{\rm m,f} &\simeq&  6.3\,{\rm MeV}\, \left(\frac{1+z}{4.512}\right)^{1/2}\xi^{-3}_{-0.12}\epsilon_{e,-0.4}^2\epsilon_{B,f,-5}^{1/2}E^{1/2}_{54.7}  t_{-0.5}^{-3/2}\cr
E^{\rm syn}_{\rm c,f}  &\simeq&  0.2\, {\rm eV}\, \left(\frac{1+z}{4.512}\right)^{-3/2}\xi^{5}_{-0.12}\,(1+x_f)^{-2}\, \epsilon_{B,f,-5}^{-3/2}\cr
&&\hspace{4.2cm}\times\,A^{-2}_{\star,-1}\, E^{1/2}_{54.7}\, t_{-0.5}^{1/2}\, \cr
F^{\rm syn}_{\rm max,f}&\simeq& 3.8\times 10^{2}\,{\rm mJy} \left(\frac{1+z}{4.512}\right)^{3/2}\xi^{-1}_{-0.12}\,\epsilon_{B,f,-5}^{1/2} \,A_{\star,-1}\,D^{-2}_{28.9}\cr
&&\hspace{4.1cm}\times\,E^{1/2}_{54.7}\,t^{-1/2}_{-0.5}\,,
\eary
}
and
{\small
\bary\label{syn_rev}
E^{\rm syn}_{\rm m,r}&\simeq& 0.2  \, {\rm keV}\, \left(\frac{1+z}{4.512}\right)^{-1/2}\xi^{-1}_{-0.12}\,\epsilon_{e,-0.4}^{2}\,\epsilon_{B,r,-0.4}^{1/2}\,\Gamma^{2}_{3}\cr
&&\hspace{4.0cm}\times A_{\star,-1}\,E^{-1/2}_{54.7}\,t_{d,-0.5}^{-1/2} \cr
E^{\rm syn}_{\rm c,r}&\simeq& 1.6\times 10^{-5} \, {\rm eV}\,  \left(\frac{1+z}{4.512}\right)^{-3/2}\xi^{5}_{-0.12}\,(1+x_r)^{-2}\cr
&&\hspace{2.9cm}\times\, \epsilon_{B,r,-0.4}^{-3/2}\,A^{-2}_{\star,-1}\,E^{1/2}_{54.7}\,t_{d,-0.5}^{1/2} \cr
F_{\rm max,r}&\simeq&  1.7\times10^8  \,{\rm \,mJy}\,  \left(\frac{1+z}{4.512}\right)^{2}\xi^{-2}_{-0.12}\,\epsilon_{B,r,-0.4}^{1/2}\,\Gamma^{-1}_{3}\,A^{1/2}_{\star,-1}\cr
&&\hspace{3.0cm}\times \,D^{-2}_{28.9}\,E_{54.7}\,t_{d,-0.5}^{-1}\,,
\eary
}
respectively,  where $x_{f/r}$ is the Compton parameter for the forward/reverse shocks.   At {\small $E^{\rm syn}=$} 100 keV,   the forward-shock synchrotron flux is in the energy range of $E^{\rm syn}_{\rm c,f} < E^{\rm syn} <E^{\rm syn}_{\rm m,f}$ (see eq. \ref{syn_for}), and then it is given by  {\small $F_{\rm \nu,f}=F^{\rm syn}_{\rm max,f} \left(E^{\rm syn}/E^{\rm syn}_{\rm c,f}\right)^{-1/2}$}  \citep{1998ApJ...497L..17S}.  Similarly, the reverse-shock synchrotron flux at {\small $E^{\rm syn}=$} 100 keV lies in the energy range of $ E^{\rm syn}_{\rm m, r} < E^{\rm syn}$ (see eq. \ref{syn_rev}). Therefore,  it can be written as {\small $F_{\rm \nu,r}=F^{\rm syn}_{\rm max, r} \left(E^{\rm syn}_{\rm m, r}/E^{\rm syn}_{\rm c, r}\right)^{-1/2}  \left(E^{\rm syn}/E^{\rm syn}_{\rm m, r}\right)^{-p/2}$} \citep{1998ApJ...497L..17S}.  Using the values reported in equations (\ref{syn_for}) and (\ref{syn_rev}), the synchrotron fluxes at forward and reverse shocks are $F_{\rm \nu,f}$ = 0.6 mJy and $F_{\rm \nu,r}$ = 38.2 mJy, respectively.  The previous result indicates that synchrotron emission from the reverse shock is dominant over that radiation originated at the forward shock.\\
The electron population submerged in an oriented magnetic field radiates by synchrotron emission. For a perfectly uniform magnetic field,  the linear polarizations of the synchrotron emission defined by power laws with  spectral indexes 1/2, (p-1)/2 and p/2 are $\Pi_{\rm max}\simeq$ 69\%, $\Pi_{\rm max}=\frac{\rm p+1}{p+\frac{7}{3}}\times 100\%\simeq$ 71\% and $\Pi_{\rm max}=\frac{\rm p+2}{p+\frac{10}{3}}\times 100\%\simeq$ 76\%, for 2.2 respectively \citep[see;][for discussion]{1999ApJ...511..852G, 2003ApJ...594L..83G}.    Taking into account  the temporal and energy range of the polarized photons detected \citep{2012ApJ...758L...1Y}, the synchrotron spectrum with the microphysical parameters  found lies in the fast-cooling regime and then, the maximum polarization degree is 69\%.  Following \cite{2003ApJ...596L..17G} for a jet with an opening angle $ 1/\Gamma \ll \theta$,  the  ordered transverse magnetic field originated at the afterglow phase can give rise in principle to observed a polarization degree as high as $\Pi_{\rm ord}\simeq$ 58\% while the position angle does not vary significantly.  This result  is in agreement with the level of polarization degree and position angle observed by the GAP instrument. Therefore, polarization properties of the early reverse shock could explain the early observations reported in \cite{2012ApJ...758L...1Y}.\\
\vspace{0.5cm}
\subsection{Origin of the subdominant prompt thermal component}
Since the subdominant prompt thermal component and the LAT light curve are present at the onset of this burst \citep{2012ApJ...757L..31A},  the main mechanisms discussed in the literature are revisited in order to interpret this subdominant prompt thermal component.  These mechanisms are: the shock break out, the hot cocoon, the synchrotron external-shock self-absorption regime and the jet photosphere.  
\subsubsection{The Shock Breakout}
Shock breakout is the interpretation given to a short X-ray burst observed from a supernova.  Shock breakouts are characterized  by peak X-ray luminosities between $10^{44} - 10^{46}\, {\rm erg/s}$ \citep{2006Natur.442.1008C, 1992ApJ...393..742E}. Since this mechanism produces a thermal emission with very low luminosity, it is not consistent with the X-ray high luminosity detected in GRB 110721A. 
\subsubsection{The Hot Cocoon}
An alternative mechanism to explain the origin for thermal emission is the hot cocoon surrounding the jet. \cite{2006ApJ...652..482P}  showed that  a few hundred-seconds  after the main emission, the cocoon emission lies in  the X-ray band and the typical radii for the emission is larger than $\geq 10^{12}$ cm. Using a relativistic expanding hot plasma cocoon, \cite{2012MNRAS.427.2950S} explained more than six bursts with black-body (BB) luminosities in the range of 10$^{47}$ to 10$^{49}$ erg/s  when these emissions occurred during the steep decay phase of the X-ray light curve.  They could not find any strong correlation between the BB properties and the prompt emission. Recently,  a hydrodynamic simulations of the hot cocoon produced when a relativistic jet goes through the  progenitor star was presented by \citet{2017arXiv170105198D}. Authors reported an isotropic cocoon luminosity of $\sim 10^{47}$ erg s$^{-1}$  which could be related with the X-ray luminosity detected during the plateau phase in a typical long-GRB afterglow.  Due to the typical values of BB luminosities and the delay times between the prompt emission/early afterglow and the thermal emission, this model cannot explain the BB luminosity and the delay time observed in GRB 110721A. 
\subsubsection{The Synchrotron External-shock Self-absorption Regime}
Other mechanism that could describe the origin of the thermal component is the synchrotron external-shock spectrum in the strong self-absorption regime \citep{2004ApJ...601L..13K,2013MNRAS.435.2520G}. In the strong absorption regime, the synchrotron self-absorption energy   ($E^{\rm syn}_{\rm a}$) is larger than synchrotron cooling energy ($E^{\rm syn}_{\rm c}$) and then,  a thermal component due to  pile-up of electrons would appear modifying the broken power-law spectrum.     Otherwise, the synchrotron spectrum  in the weak-absorption regime ({\small $E^{\rm syn}_{\rm a} <  E^{\rm syn}_{\rm c}$})  is not altered by the self-absorption process.\\
Using the observable quantities of GRB 110721A and requiring the microphysical parameter values found,  the synchrotron self-absorption energies using z=3.512 for forward and reverse shocks are
\bary\label{Eaf}
E^{\rm syn}_{\rm a,f} &\simeq&  1.1\times 10^{-3}\,{\rm eV}\, \left(\frac{1+z}{4.512}\right)^{-2/5}\xi^{-6/5}_{-0.12}\,\epsilon_{e,-0.4}^{-1}\,\epsilon_{B,f,-5}^{1/5}\cr
&&\hspace{3.3cm}\times\,A^{6/5}_{\star,-1}\,E^{-2/5}_{54.7}\,  t_{-0.5}^{-3/5}
\eary
and
\bary\label{Ear}
E^{\rm syn}_{\rm a,r}&\simeq& 3.5\times 10^{-10} \, {\rm eV}\, \left(\frac{1+z}{4.512}\right)^{-7/5}\xi^{4/5}_{-0.12}\,\epsilon_{e,-0.4}^{-1}\,\cr
&&\hspace{1.0cm}\times \epsilon_{B,r,-0.4}^{1/5}\,\Gamma^{2}_{3}\,A_{\star,-1}^{11/5}\,E^{-7/5}_{54.7}\,t_{d,-0.5}^{2/5}
\eary
respectively.  By comparing the equations  (\ref{syn_for}) - (\ref{Ear}), it can be seen that the self-aborption energies from the forward and reverse shocks lie in the weak-absorption regime, thus discarding this mechanism as possible origin of the subdominant prompt  thermal component. 
\subsubsection{The Jet Photosphere}
The thermal photospheric component is related with the optically thick plasma of a relativistic jet \citep{2011ApJ...727L..33G, 2013ApJ...770...32G, 2015ApJ...814...10G}.  The thermal emission from the photosphere emerges when the optical depth of this fireball plasma becomes unity. \cite{2013ApJ...771...15F} presented a search for thermal emission in the early X-ray afterglows.  They identified  a thermal component in eight bursts, determining very large luminosities (10$^{48}$ to 10$^{52}$ erg/s), photospheric radii ($\sim$ 10$^{13}$ - 10$^{15.5}$ cm)  and  temperatures for many of them.  They proposed that the thermal component in coincidence with the early $\gamma$-ray /X-ray afterglow could be modelled as late photospheric emission from the jet.   They even claimed that this model could account for the thermal component present in GRB 110721A.   \cite{2013MNRAS.433.2739I} studied GRB 110721A in the framework of photospheric emission. They found that the bulk Lorentz factor decreases monotonically with time from $\sim$ 1000 to $\sim$ 150.  Assuming a black hole mass of 10 M$_\odot$, authors found that jet was  moderately magnetized with a magnetization parameter of $\sigma\sim$ 0.8.\\
If the dissipative effects below the photosphere take place (e.g. magnetic reconnection, shocks, etc), a copious pair formation dominates the photospheric opacity.  In this case, the pair production induces a new photosphere farther out than the common baryonic photosphere, thus delaying the thermal emission \citep{2005ApJ...628..847R, 2011ApJ...726...90Z}.\\
\cite{2008ApJ...682..463P} showed that the photospheric radius strongly depends on the angle to the line of site. Thermal emission can be observed as long as tens of seconds following the decline of the central engine.  Observation of the thermal emission at early times when it is observed along the line of site provides an unequivocal measurement of the temperature and photon flux. Otherwise,  thermal emission detected at late times when the emission is off-axis could be observed and then could be relevant in the early afterglow phase, similar to the high-latitude emission discussed in the context of GRB afterglow \citep{1996ApJ...473..998F, 2000ApJ...541L..51K}. \\
\\
Considering that X-ray high luminosity of the subdominant prompt thermal component can be overlapped with the early afterglow phase, and also the values obtained in the early-afterglow external shock model similar to those reported by \cite{2013MNRAS.433.2739I}:  i)  The jet moderately magnetized  with a value of magnetization parameter $\sigma\sim$ 0.3-0.5 and  ii) the decrease of the value of bulk Lorentz factor $\Gamma\sim$ 1000, therefore the subdominant prompt thermal component  detected in GRB 110721A  favors the late photospheric emission.
%
%
%
\section{Conclusions}
%
%
Analyzing the LAT data around the reported position of GRB 110721A with the most recent software SICENCETOOLS version v10r0p5 and reprocessing them with ``Pass 8'' extended, spacecraft data and the instrument response functions P8R2\_TRANSIENT020\_V6, the LAT light curve  above 100 MeV and the photon index of GRB 110721A is reported.  The LAT light curve presents a short-lived component peaking at $\sim$ 0.25 - 0.35 s and also a long-lived emission lasting hundreds of seconds.  We have proposed that this light curve can have as origin those radiative processes generated in external shocks; the long-lived component by synchrotron radiation from forward shock and the short-lived emission by SSC emission from reverse shock. Additionally,  we have found that the propagating outflow into a stellar wind evolved in the thick-shell case and must be moderately magnetized. The early high-energy photons ($>$100 MeV) detected by LAT support the idea that these have as origin the external-shock region instead of internal-shock region as have been claimed by some authors \citep{2011MNRAS.415...77M, 2011ApJ...733...22H,  2011ApJ...730....1L}.\\


We discuss the origin and the implications of the early-afterglow external shock model on the polarization observations, the extreme peak energy at 15$\pm$2 MeV and the subdominant prompt thermal component. 
\begin{enumerate}
\item Based on the fact that photospheric and internal shock emission can hardly reproduce  the high degree of linear polarization with a stable position angle and the LAT emission described in the early afterglow model,  we show that reverse shock could explain the polarization properties observed in this burst. Using the condition of an opening angle $ 1/\Gamma \ll \theta$ for the jet   \citep{2003ApJ...596L..17G}, the ordered transverse magnetic field originated at the afterglow phase could in principle give rise to  a polarization degree as high as $\Pi_{\rm ord}\simeq$ 58\% while the position angle does not vary significantly, as observed in GRB 110721A.  It is worth noting that with the microphysical parameters found  after describing the LAT emission, the  synchrotron emission obtained from the reverse-shock region is dominant over the synchrotron flux from the forward shock.

\item Taking into account the peak exhibited in the LAT emission as the beginning of the deceleration of the jet, and the values of the magnetization parameters found after fitting the LAT light curve,  it can be seen that the most suitable process to describe the extreme energy peak at 15 $\pm$ 2 is the ICMART model  \citep{2011ApJ...726...90Z}.  Studying the magnetization degree in the jet, and the evolution and the spectral features of the extreme peak energy,  \cite{2012ApJ...758L..34Z},  \cite{2013MNRAS.433.2739I} and \cite{2015ApJ...801..103G} similarly found that the best candidate to explain this atypical peak was the  ICMART process. 

\item Several mechanisms were discussed in order to interpret the subdominant prompt thermal component. Taking into consideration the degree of magnetization and the decreased value of bulk Lorentz factor as found after modeling the LAT emission and also  
the high luminosity of this  thermal component, the late photospheric emission model is favored to describe the subdominant prompt thermal component.  

\end{enumerate}
 The early-afterglow model used to describe the multiwavelength observations in GRB 110721A  suggests that outflow is moderately magnetised $\sigma\sim$ 0.3 - 0.5.  This value is consistent with the fact that the reverse shock is strong and then the short-lived LAT emission could be successfully  described \citep{2005ApJ...628..315Z}.  A similar value of magnetization degree ($\sigma\sim$ 0.8) was found by \cite{2013MNRAS.433.2739I} after describing  the thermal photospheric data.   This magnetization is possible provided that the magnetic acceleration is inefficient \citep[e.g. see,][]{2012ApJ...761L..18V}.\\ 
Regarding that the ejecta propagating into the stellar wind is quickly decelerated, at $\sim 0.3$ s, and using the typical value of stellar wind density of $A= (5.0\times 10^{10})\,{\rm g/cm}$ \citep{2000ApJ...536..195C,  2014Sci...343...42A,2013ApJ...763...71A,  2016ApJ...818..190F,2015ApJ...804..105F}, then the bulk Lorentz factor is about $\Gamma\sim$ 1000 \citep{2012ApJ...755...12V, 2013ApJ...763...71A, 2014Sci...343...42A}.  This value is equal  to that found by \cite{2013MNRAS.433.2739I} after describing the subdominant prompt thermal  data as photospheric emission. \\
%
%
The early-afterglow external shock model proposed to explain the high-energy emission in GRB110721A indicates that the outflow carries a meaningful magnetic field.  Similar results have been found in the most luminous LAT-detected bursts such as  GRB090902B \citep{2009ApJ...706L.138A},   GRB110731A \citep{2013ApJ...763...71A, 2015ApJ...804..105F},  GRB130427A \citep{2014Sci...343...42A,2016ApJ...818..190F},   GRB160625B \citep{2016arXiv161203089Z, 2017arXiv170509311F}, among others. What makes it unique is the large amount of Poynting flux that must have been dissipated during the prompt emission phase.
\section*{Acknowledgements}
We thank Dimitrios Giannios, Fabio de Colle, Bing Zhang, Anatoly Spitovsky and Simone Dichiara for useful discussions.  NF  acknowledges  financial  support from UNAM-DGAPA-PAPIIT through grant IA102917 and  WHL  through  grant  100317.    PV  thanks  Fermi  grant NNM11AA01A and partial support from OTKA NN 111016 grant.
%
%

%
\clearpage

\end{document}